\documentstyle[aps,prc,epsf]{revtex}
\begin{document}
\title{Long-range correlations in finite nuclei: comparison of two 
self-consistent treatments}
\author{Y. Dewulf, D. Van Neck, L. Van Daele and  M. Waroquier} 
\address{Laboratory for Theoretical Physics,
Proeftuinstraat 86, B-9000 Gent, Belgium}
\draft
\maketitle
\vspace*{.5cm}
\begin{abstract}
Long-range correlations, which are partially responsible for the observed
fragmentation and depletion of low-lying single-particle strength, are studied
in the Green's function formalism. The self-energy is expanded up to second order
in the residual interaction. We compare two methods of implementing
self-consistency in the solution of the Dyson equation beyond Hartree-Fock,
 for the case of the $^{16}$O nucleus.
It is found that the energy-bin method and the BAGEL method lead to 
globally equivalent results. In both methods the final single-particle strength
functions are characterized by exponential tails at energies far from
the Fermi level.
\end{abstract}
\vspace*{.5cm}

\section{Introduction}

The effect of nucleon-nucleon (NN) correlations on single-particle motion
in finite nuclei has been the subject of many experimental and theoretical 
investigations \cite{deWi,Lapi,Maha}. 
The most striking experimental fact is the large 
suppression (by a factor of about 0.65 on average \cite{Lapi}) of 
the low-lying
hole strength seen in $(e,e'p)$ reactions, compared to the values in 
the independent particle model. Both short-range (SRC) and long-range (LRC)  
correlations seem to play a dominant role in the observed depletion of 
hole strength. 
In order to explain this, one 
ususally discriminates between the effect of short-range
 (SRC) and long-range (LRC) correlations. Most recent calculations 
\cite{Piep,PRC49} now indicate  that SRC, which originate from the strong 
short-range and tensor 
components of any realistic  nuclear force, remove about 10\% of the 
strength in the
quasi-hole region. After accounting for the effects of SRC, which  should be 
rather $A$-independent,  the dynamics in the low-energy region 
is governed by a well-behaved effective interaction. The additional 
configuration mixing and coupling to collective states are known 
 to induce a further depletion and strong 
fragmentation of the low-lying single-particle (s.p.) strength 
\cite{VanN1,VanN2,VanN3,Sluy} . 
These LRC effects are dependent on the shell structure of real nuclei 
and cannot be extracted from nuclear matter calculations, but should be
calculated explicitely for each specific nucleus.

Recently interest has arisen into {\em ab initio} calculations of LRC 
starting from
realistic NN interactions \cite{NP555,PL306,NP581,NP604,Bran,Rijs,Geur}. 
The effects of the hard core are 
handled by the construction of an effective G-matrix interaction 
suitable for a limited model
space of active nucleons in s.p.\ states around the Fermi level. 

Especially 
self-consistent Green's function methods \cite{VanN2,VanN3,PL306} have 
received current interest. 
In this context,  self-consistency means that the 
internal lines in the self-energy diagrams should obey the Dyson equation
themselves. According to fundamental theorems \cite{Baym}, 
the self-consistency
requirement is necessary for the theory to obey basic conservation laws, such
as conservation of energy and particle number.

The implementation of the self-consistency 
requirement in the description of finite nuclei, however, is not trivial. 
One of the problems is related  to the strong non-linearity of the 
resulting Dyson equation, which necessitates an iterative solution.
The self-energy expanded to second order in the residual 
interaction, {\em e.g.}, contains three exact propagators. If one works in 
a discrete model space this means that, starting from 
a representation of the Green's function in terms of a number $N$ of discrete 
poles, the  Green's function in the next iteration has a number of poles in
 the order of $N^3$, 
and the dimensionality quickly gets out of hand.  

Two methods have been proposed to solve this problem. 
Van Neck {\em et al.} \cite{VanN2,VanN3}  
discretized the energy axis and represented the spectral strength of the 
Green's function and self-energy in equidistant energy bins. This approach 
will be referred to as the BIN method. In the 
BAGEL (BAsis GEnerated by Lanczos, see \cite{NP482}) approach  
M\"{u}ther {\em et al.} 
represented \cite{PL306} the Green's function in terms of a few characteristic 
poles, chosen in such a way that the lowest order moments of the exact 
distribution are reproduced. 

It is not {\em a priori} clear that both methods give identical 
results when iterated to convergence. 
In particular, the weak point in the 
BAGEL method seems to be the fact that the lowest order moments of the 
{\em global} spectroscopic distribution are reproduced. It is not guaranteed
that the BAGEL representation yields in all cases an accurate representation of the 
strength distribution in the $(A-1)$ and $(A+1)$-particle system 
separately. The construction of the self-energy in the next-order iteration
is, however, dependent on this separation into $(A-1)$ and $(A+1)$ strength.
The BIN method, on the other hand, reproduces all moments simultaneously, but only
on the average due to the finite width of the employed energy bins.

It is the purpose of this letter 
to compare the results obtained by both methods for the case of the 
$^{16}$O nucleus. Apart from global quantities
such as the binding energy, occupation probabilities and spectroscopic factors
of the main fragments, the spectral distribution in the $(A-1)$ and $(A+1)$-particle
system are compared in more detail by calculating their higher order moments. 

\section{Formalism}

For the purpose of a clear comparison between the two methods we make some simplifying 
assumptions: {\em (i)} We assume that the Green's function  and the self-energy are
 both diagonal in 
the adopted spherical and discrete s.p.\ basis, {\em i.e.}  we neglect major shell-mixing.
This is justified since the energy spacing between major shells is sufficiently large to
suppress mixing between s.p.\ states with same parity and angular momentum but
 different radial
quantum number. The influence of major shell-mixing  in the 
self-energy was investigated in \cite{VanN3} for the case of $^{208}Pb$, and found to be small.
{\em (ii)} We also ignore the residual interaction between the 2p1h and 2h1p configurations.
This approximation was studied in \cite{NP581,NP604} and will change 
the results only by a few percent.

The second order self-energy $\Sigma$ is now represented diagrammatically in Fig.~1, and the 
Dyson equation to be solved takes the form
\begin{equation}
G_{\alpha}(\omega)=G_{\alpha}^{HF}(\omega)+G_{\alpha}^{HF}(\omega)
\Sigma_{\alpha}(\omega)
G_{\alpha}(\omega)\;,
~\label{eq:Dyson}
\end{equation}
with $\alpha$ the spherical s.p.\ label $nljm$,  $G$ the exact (second order) Green's function
 and  $G^{HF}$ the Hartree-Fock Green's function.

The self-consistency of our approach implies that we must use the solution 
$G_{\alpha}(\omega)$ itself in the evaluation of the self-energy 
$\Sigma_{\alpha}(\omega)$,
\begin{eqnarray}
\Sigma_{\alpha}(\omega)&=&\frac{1}{2}(\frac{1}{2\pi})^{2}
\sum_{\gamma_1 \gamma_2 \gamma_3}
|\langle \alpha\gamma_3 |\hat{V}|\gamma_1 \gamma_2 \rangle_{as}|^{2}
\nonumber\\
&&\times\int d\omega_{1}d\omega_{2}G_{\gamma_1}(\omega_{1})
G_{\gamma_2}(\omega_{2})G_{\gamma_3}(\omega_1 +\omega_2 -\omega)
\;.            
\label{eq:self-en}
\end{eqnarray}
In Eq.(\ref{eq:self-en}) 
$\langle \alpha\gamma_3 |\hat{V}|\gamma_1 \gamma_2 \rangle_{as}$ stands for the
matrix elements of the effective interaction between 
antisymmetrized two-nucleon states.
Eq.(\ref{eq:Dyson}) with a self-energy given by eq.(\ref{eq:self-en}) 
is highly non-linear in the propagator and must be solved by iteration. 
Starting from a certain approximation for the Green's 
function, the self-energy can be evaluated using eq.(\ref{eq:self-en}).
In the next iteration this new estimate for the self-energy is used as input 
in solving the Dyson equation (\ref{eq:Dyson}) to get a new 
approximation for the Green's function. This iteration procedure is repeated
until sufficient convergence is reached.
Here, we encounter the dimensionality problem mentioned earlier: 
if $\Sigma_{\alpha}^{(n)} $ has N poles,  
$G_{\alpha}^{(n+1)}$ will have N+1 poles, and  $\Sigma_{\alpha}^{(n+1)}$ 
will have a number of poles 
in the order of $N^{3}$, 
{\em i.e.}  the number of poles is roughly cubed after each iteration.  
Note that this is a byproduct of the discretization of the s.p. continuum; in 
reality the propagator and self-energy will have, in addition to the discrete poles, branch-cuts
representing the continuum states in the $(A-1)$ and $(A+1)$-particle systems.
The BIN and BAGEL methods are essentially two different ways of limiting  the number of poles after 
each iteration. For the sake of completeness we briefly summarize the two 
methods below. 
A more extensive treatment can be found in \cite{VanN2} and \cite{NP581}. 

In the BIN method the energy axis is divided into a  large number $M_B$ of 
equidistant energy-bins (typically a few thousand). 
This means that in each iteration step the self-energy and Green's function are represented by their 
residues at a number of fixed poles, which makes it possible to develop 
an iterative scheme for solving eq.(\ref{eq:Dyson})
with a self-energy given by eq.(\ref{eq:self-en}). The Green's function 
after $n$ iterations will then be 
given by:
\begin{equation}
G_{\alpha}^{\,(n)}(\omega)=\sum_{k}\frac{f_{\alpha,k}^{(n)}}
{(\omega-E_{k}+i\eta)}
+\sum_{k}\frac{ b_{\alpha,k}^{(n)}}{(\omega+E_{k}-i\eta)}\;,
\end{equation}
with $E_{k}=k\Delta$, for $k=1,\dots,M_B$. 
Using this propagator we can calculate the self-energy 
$\Sigma_{\alpha}^{(n)}$ ,
\begin{equation}
\Sigma_{\alpha}^{(n)}(\omega)=\sum_{k}\frac{\sigma_{\alpha,k}^{(+)(n)}}
{\omega-E_{k}+i\eta}
+\sum_{k}\frac{\sigma_{\alpha,k}^{(-)(n)}}{\omega+E_{k}-i\eta}\;,
\end{equation}
with
\begin{eqnarray}
\sigma_{\alpha,k}^{(+)(n)}&=&\frac{1}{2}
\sum_{\stackrel{\scriptstyle 
k_{1}k_{2}k_{3} }{\scriptscriptstyle (k=k_{1}+k_{2}+k_{3})}}
\sum_{\gamma_1 \gamma_2 \gamma_3}
|\langle \alpha\gamma_3 |\hat{V}|\gamma_1 \gamma_2 \rangle_{as}|^{2}
f_{\gamma_1 ,k_{1}}^{(n)}f_{\gamma_{2},k_{2}}^{(n)}b_{\gamma_{3},
k_{3}}^{(n)}\;,
\label{eq:self1}\\
\sigma_{\alpha,k}^{(-)(n)}&=&\frac{1}{2}
\sum_{\stackrel{\scriptstyle 
k_{1}k_{2}k_{3} }{\scriptscriptstyle (k=k_{1}+k_{2}+k_{3})}}
\sum_{\gamma_1 \gamma_2 \gamma_3}
|\langle \alpha\gamma_3 |\hat{V}|\gamma_1 \gamma_2 \rangle_{as}|^{2}
b_{\gamma_{1},k_{1}}^{(n)}b_{\gamma_{2},k_{2}}^{(n)}
f_{\gamma_{3},k_{3}}^{(n)}\;.
\label{eq:self2} 
\end{eqnarray}
Note that the triple summation over the equidistant binpoints in  
eqs.(\ref{eq:self1}-\ref{eq:self2}) for the self-energy is in fact 
a triple discrete convolution, which 
can be performed very efficiently using Fast Fourier Transform techniques.

The next approximation for the Green's function is then given by,
\begin{equation}
G_{\alpha}^{(n+1)}(\omega)=\frac{1}{\omega-\epsilon_{\alpha}^{HF}
-\Sigma_{\alpha}^{(n)}(\omega)}\;,
\label{eq:GafoS2}
\end{equation}
with $\epsilon^{HF}_\alpha $ the Hartree-Fock s.p. energy of s.p.\ state 
$\alpha$.
The poles of $G_{\alpha}^{(n+1)}$ are 
located at the solutions  $\omega_i$ of the equation  
\begin{equation}
	\omega_i =\epsilon_{\alpha}^{HF}+\Sigma_{\alpha}^{(n)}(\omega_i )\;,
\label{eq:find-poles}
\end{equation}
and carry a strength $s_i$ equal to 
\begin{equation}
s_i =\frac{1}{1-\frac{d}{d\omega}(\Sigma_{\alpha}^{(n)}(\omega))
|_{\omega=\omega_i }} .
\end{equation}
This strength is assigned to the binpoint closest to $\omega_i$. 
The error coming from this averaging procedure can be made sufficiently small by choosing 
small widths $\Delta$ for the energy-bins.

The concept of the BAGEL approach is based on a representation of the 
Green's function in terms of a small number of 
poles, with judiciously chosen positions and residues, in such a way that 
the lowest order moments of 
the s.p. strength distribution  are reproduced. 
Assume that the Green's function $G_{\alpha}^{(n)}(\omega)$ for each 
s.p.\  state has $N$ poles.
Then the corresponding self-energy $\Sigma_{\alpha}^{(n)}(\omega)$ will have 
a larger number $D$ of poles (in the order of $N^3$). In the next iteration 
step the ''true'' Green's function,
\begin{equation}
G_{\alpha,true}^{(n+1)}(\omega)=\frac{1}{\omega-\epsilon_{\alpha}^{HF}
-\Sigma_{\alpha}^{(n)}(\omega)}
=\sum_{i=1}^{D+1}\frac{|X_{\alpha,i}^{(n+1)}|^{2}}
{\omega-\omega_{\alpha,i}^{(n+1)}} \;,
\label{eq:bagit}
\end{equation}
will have $D+1$ poles. 
The moments $m_p$ of the corresponding strength distribution are defined 
according to 
\begin{equation}
\label{m_p}
m_{p}=\sum_{i=1}^{D+1}(\omega_{\alpha,i}^{(n+1)})^{p}
(X_{\alpha,i}^{(n+1)})^{2}.
\end{equation}
It is possible to determine a smaller set of poles, with positions and 
residues different from 
the true ones, but reproducing exactly some of the lowest-order moments 
$m_p$ given in Eq.(\ref{m_p}). 
The BAGEL procedure determines this smaller set by a truncated Lanczos 
tridiagonalisation of the Hamiltonian submatrices corresponding to the 
$1p\oplus 2p1h$ space
and $1h\oplus 2h1p$ space. The true Green's function is then replaced by 
$G_{\alpha}^{(n+1)}$ 
containing this smaller set of poles.

Various BAGEL approximations are possible, depending on the number of 
Lanczos vectors
constructed in the two subspaces. In the so-called BAGEL(M,M) approach, 
used in this work, the approximated Green's function has $(2M+1)$ poles and at 
each iteration step the moments $m_p$, for $p=0,\dots,2M+1$, of the total 
true strength distribution will be reproduced. 

\section{Numerical results}

The construction of the effective interaction and choice of the s.p.\ basis
are analogous to those in \cite{NP581}, to which we refer for more details.

We start from a realistic interaction, the Bonn-C potential \cite{Mach}.
The effective interaction $\hat{G}$ in $^{16}$O is derived by solving the 
Bethe-Goldstone
equation, with a Pauli operator consistent with a discrete model space
of harmonic oscillator states (with frequency $\hbar\omega_0 = 13.27 MeV$) 
up to and including the $pf$ shell. A value of $\omega=-30 MeV$ is taken 
for the starting energy. 

The basis of oscillator states can be  identified with the Hartree-Fock basis, 
after subtracting from the original effective interaction a correction 
term \cite{NP581},
\begin{equation}
\hat{V}=\hat{G}-\sum_{\alpha\neq\beta}(t_{\alpha\beta}+U_{\alpha\beta})
c^+_{\alpha}c_{\beta},
\end{equation}
in which $t_{\alpha\beta}$ stands for the matrix element of the kinetic 
energy operator
 and $U_{\alpha\beta}=\sum_{h}\langle\alpha h |G|\beta h\rangle_{as}$.
The Hartree-Fock single-particle energies are then obtained as
 $\epsilon^{HF}_{\alpha}=t_{\alpha\alpha}+U_{\alpha\alpha}$,
 and are listed in Table~1.

The second order Dyson-equation is then solved self-consistently within 
 the BAGEL and BIN methods. In both methods we use renormalized Hartree-Fock
(RHF) s.p.\ energies in order to take the newly generated contributions to 
the first order self-consistent self-energy into account, {\em i.e.}
after each iteration the s.p.\ energies are changed according to 
\begin{equation}
\epsilon_{\alpha}^{RHF}=t_{\alpha\alpha}
+\frac{1}{2}\sum_{\beta}n_{\beta}
\langle\alpha\beta|\hat{V}|\alpha\beta\rangle_{as}\;,
\end{equation}
with $n_{\beta}$ the new occupation of the s.p.\ state $\beta$.

After studying the dependence of the BAGEL(M,M) approach on the number of 
basis vectors retained in the truncated subspaces we find that the results 
are converged for $M\geq 6$. We will display the results for the BAGEL(6,6) 
approximation. In the  
BIN method we used the following choice of parameters: a bin width 
$\Delta=0.05 MeV$ and a number of bins $M_B = 5460$.  

After the first iteration the number of particles has a small (0.5\%) excess.  
When convergence is reached the restoration of particle number conservation, 
typical for a self-consistent solution, is observed in both approaches. 
We also find that after one iteration the effect of correlations is 
somewhat overestimated, in agreement with the results in \cite{VanN2,PL306}.

In Table~1 the RHF energies and occupation probabilites of all s.p.\ states
are shown. These are in complete agreement for the two methods. The same 
agreement holds for the mean removal energies (not shown in Table~1), and 
hence also for the total binding energy per nucleon, calculated via the 
Koltun sum rule, which is listed on the last row in Table~1.     

In the last columns of Table~1 we compare the position and strength of the 
main fragment in both methods. The valence shells 
($1p$, $2s$, $1d$) have a clear quasi-particle/hole behaviour, with one pole 
carrying  about 90\% of the total s.p.\ strength. The positions and strengths 
of these states are equivalent in both approaches. 
A more detailed comparison reveals that the strengths of the largest peaks 
are systematically higher in the BAGEL approach than in the BIN method,
 an effect that becomes more pronounced away from the Fermi level. This is 
due to the fact that in the BAGEL approach, because of the limited number of 
poles, the main peaks need to carry some additional strength corresponding to 
states which in the BIN method are really separate states, {\em i.e.} 
the BIN method allows for more fragmentation. 

This is seen more clearly in Fig.~2, where the total spectral functions of 
the $1s_{1/2}$ and $1d_{5/2}$ states are shown. Apart from this enhanced
fragmentation into discrete states for energies up to about 30 MeV above and  
below the Fermi level, one also observes for larger energies the appearance
of a quasi-continuous distribution in the BIN results. This part of the 
distribution is of course also represented in the BAGEL method, in terms of
a few well-separated states. 

A very surprising feature, appearing in both the 
BAGEL and BIN methods, is the nearly perfect exponential decay 
of the spectral function at these large excitation
energies. The decay constant has the same value in both methods. To within a 
few percent, this value is independent of the s.p.\ state, and the same 
for the strength tails in the $(A+1)$ and $(A-1)$ system.
Since the determination of the Green's function in each iteration step
only involves rational functions, the exponential behaviour must be built up
by the iterative procedure itself, {\em i.e.} it is a genuine result 
of the requirement of self-consistency for the Green's function.  
We do not know whether it is also a general feature of the exact Green's 
function, without two limitations of the present model:
{\em (i)} The G-matrix is used here as a static effective interaction 
\cite{NP581}. In principle the energy-dependence of the G-matrix must be 
taken into account for large excitation energies in the $(A\pm 1)$ system, 
and this may affect the behaviour of the spectral function in this energy 
domain.  
{\em (ii)} The s.p.\ continuum is discretized, and is represented in
terms of a few unoccupied s.p.\ states, {\em i.e.} we work in a truncated 
model space. It is unclear whether the exponential
decay of the iterated spectral function  would persist if already in the 
first iteration step, on the 2p1h-2h1p level, the self-energy strength is  
spread out over all energies (as will be the case with an exact treatment 
of the continuum). 

The exponential localization of strength distributions in a finite
model space was also found in a recent investigation \cite{Fraz}, where the
statistical properties of the $J^{\pi}T=0^+ 0$ states for 16 active nucleons
in the $sd$ shell were studied. This feature seems to be poorly understood 
at present. Note that self-consistent Green's function methods, which include
many-body damping in a natural way \cite{VanN2}, can also provide a dynamical 
framework for the process of stochastization in complex wavefunctions. It
should {\em e.g.} be possible to derive the localization length directly
in our model. Work along these lines is in progress.
 
In Fig.~3 the detailed behaviour of the $(A-1)$ part of the
$l=1$ spectral function is shown in the energy region above the quasi-hole
peaks. The spectral functions have been folded with a Lorentzian having a 
full width of 1 MeV. 
In the BAGEL approach the spectral strength is characterized by two peaks of 
about the same energy, whereas in the BIN approximation
a broader distribution appears with a natural spreading width. 

The BAGEL(M,M) method reproduces, at each iteration step, the moments of 
the total strength distribution up to order $2M+1$. This property, however,
is not conserved through succesive iterations, since the self-energy depends
on the separation of the spectral function into its $(A-1)$ and $(A+1)$ 
parts. So strictly speaking there is no exact relationship between the moments
of the true (converged) spectral function and its BAGEL approximation, though
in practice the BAGEL property is so restrictive that also the 
separate $(A-1)$ and $(A+1)$ parts will be, in most cases, well
described in terms of a few BAGEL states. This holds especially for
fairly weakly correlated Fermi systems, where the spectral function is
dominated by either the $(A-1)$ or $(A+1)$ part, since in that case the
reproduction of the moments of the total spectral function and of those of 
its dominant part is practically coinciding.
 
In order to study this point it is interesting to compare the moments of the 
BAGEL spectral functions with those obtained with the BIN method, since in the
 latter approach the errors due to averaging are expected to be spread evenly
over the entire energy range. In Table~2 we list the relative differences
between the central moments of the $(A-1)$, $(A+1)$ and total 
s.p.\ strength distribution, obtained in both methods. 
The agreement for the total distribution is impressive, with deviations 
of less than 1\% up to the sixth moment. The same holds for the moments of the 
dominant part of the distribution (the $(A-1)$ part for the hole states, the 
$(A+1)$ part for the particle states), whereas  the deviations for the 
non-dominant part are somewhat larger. Only for the highest moments the 
two methods start to deviate substantially. 

For future applications that go beyond the second-order level it is
also of interest to compare the numerical workload required by both methods. 
Of course this workload depends crucially on the number of Lanczos vectors
in BAGEL or on the number of energy bins in the BIN method. 
For the present choice of parameters the computational effort is of 
similar magnitude for both methods, if Fast Fourier Transform is used in 
the BIN method to evaluate the discrete convolutions in 
eqs.(\ref{eq:self1}-\ref{eq:self2}) (this was not implemented in earlier 
calculations \cite{VanN2,VanN3}). The storage requirement for the Green's 
function, however, is considerably less in the BAGEL method (in which the 
Green's function is represented by a small number of poles and residues) 
than in the BIN method (in which the value of the spectral function at 
each energy binpoint must be stored).

\section{Summary}

We compared two theoretical methods, that aim at implementing the 
important requirement of self-consistency in the framework of Green's  
function theory, by calculating the single-particle spectral functions 
in $^{16}$O. We found that they yield equivalent results for all global 
quantities that depend on the lowest order moments, such as occupation 
probabilities, mean removal energies and total binding energy. 
The prediction for the strength and position of the valence state 
quasi-particle/hole peak are also in close agreement.  
The BIN method allows for more fragmentation resulting in structures 
at intermediate energies that have a natural spreading width.    

In both methods the spectral functions exhibit an exponential decay  
at large excitation energies in the $(A-1)$ and $(A+1)$ system.
It is likely that this is a result of the requirement of self-consistency.

\subsubsection*{Acknowledgments}

This work is supported by the Fund for Scientific Research - Flanders (FWO).
We are grateful to H.\ M\"{u}ther for providing us with the G-matrix elements
used in this work.

\newpage

\begin{figure}[Fig.1]
\epsfbox{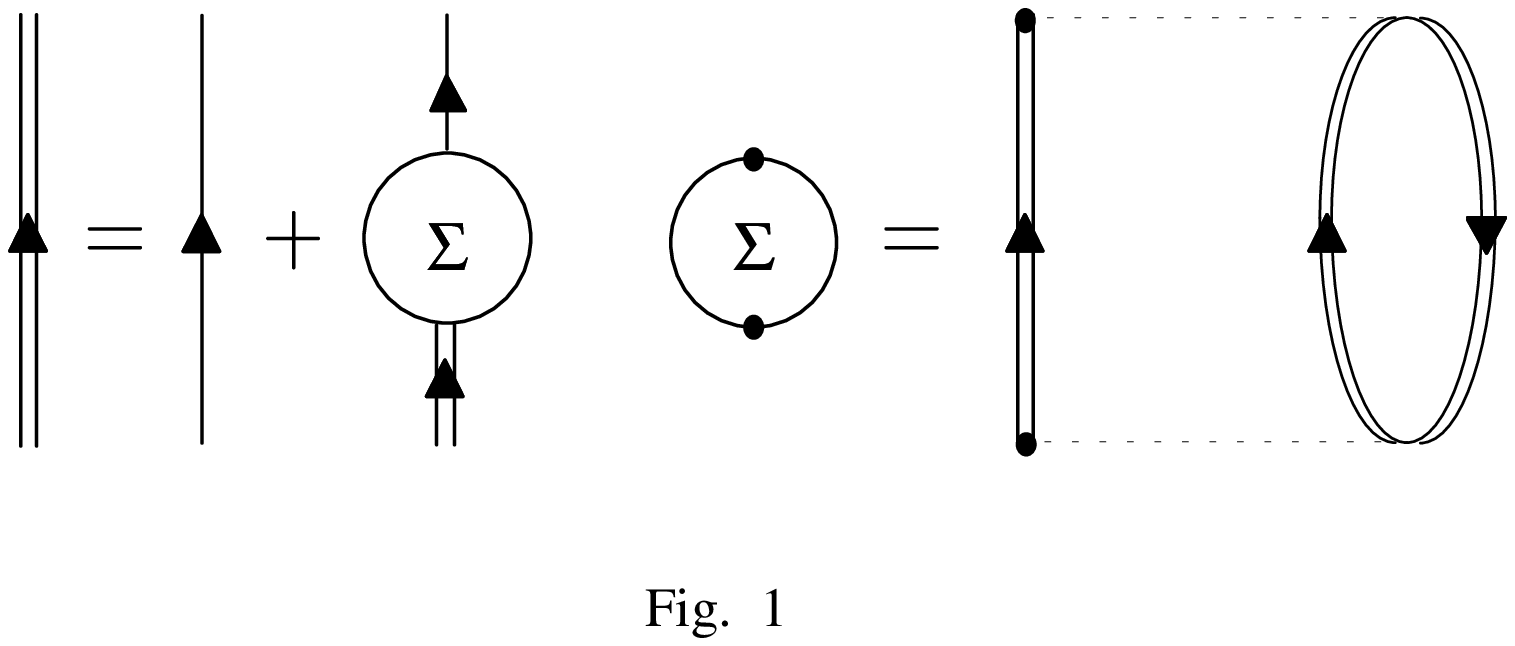}
\caption{Diagrammatical representation of second order self-consistent Dyson 
equation}
\end{figure}
\newpage
\begin{figure}[Fig.2]
\epsfysize=21cm
\epsfbox{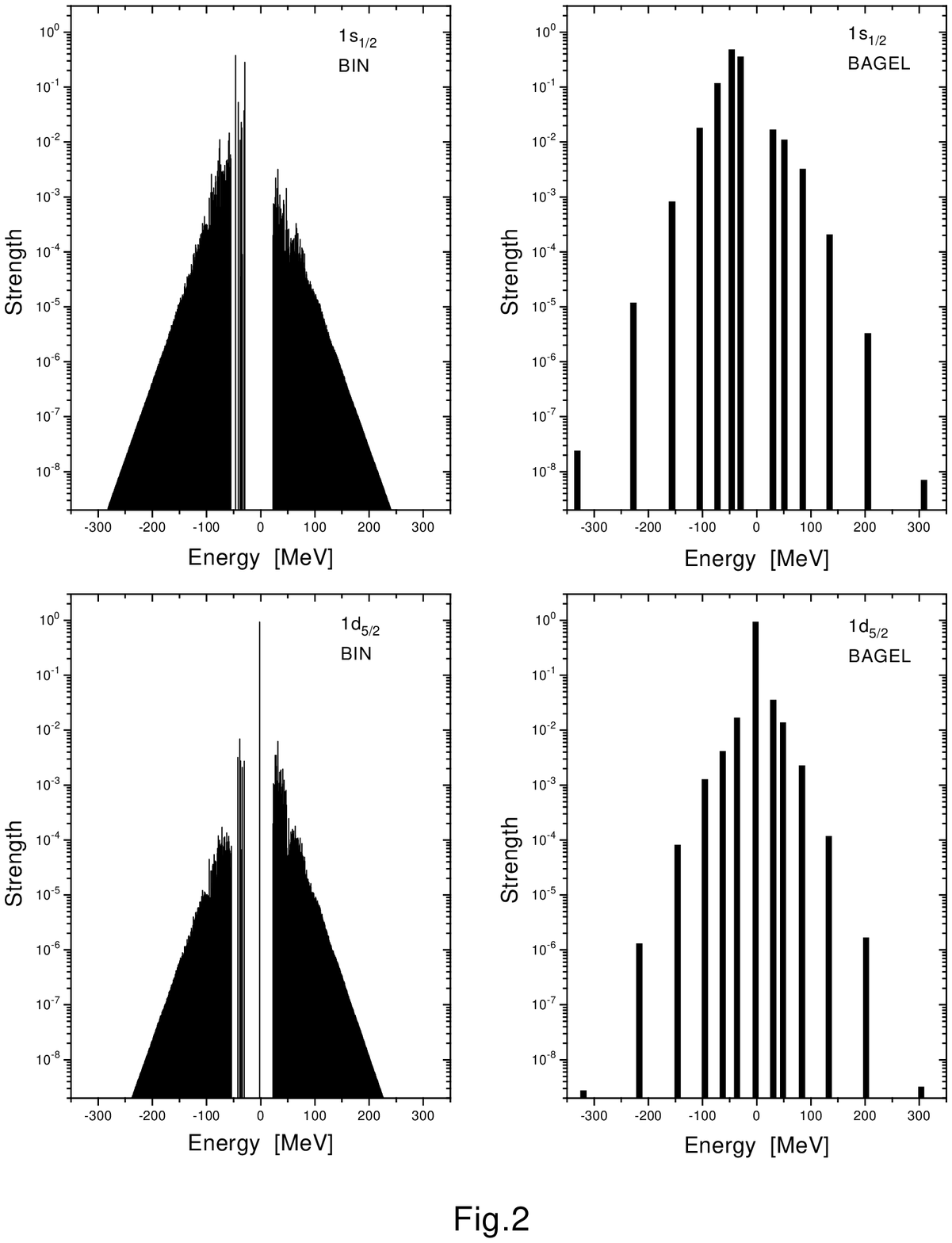}
\caption{Total (discrete) spectral function for the $1s_{1/2}$ and $1d_{5/2}$ 
s.p.\ states in $^{16}$O, obtained
with BIN and BAGEL method, after convergence. For practical reasons 
 the strength has been resummed in intervals of 0.25 MeV for the BIN result.
}
\end{figure}
\newpage
\begin{figure}[Fig.3]
\epsfbox{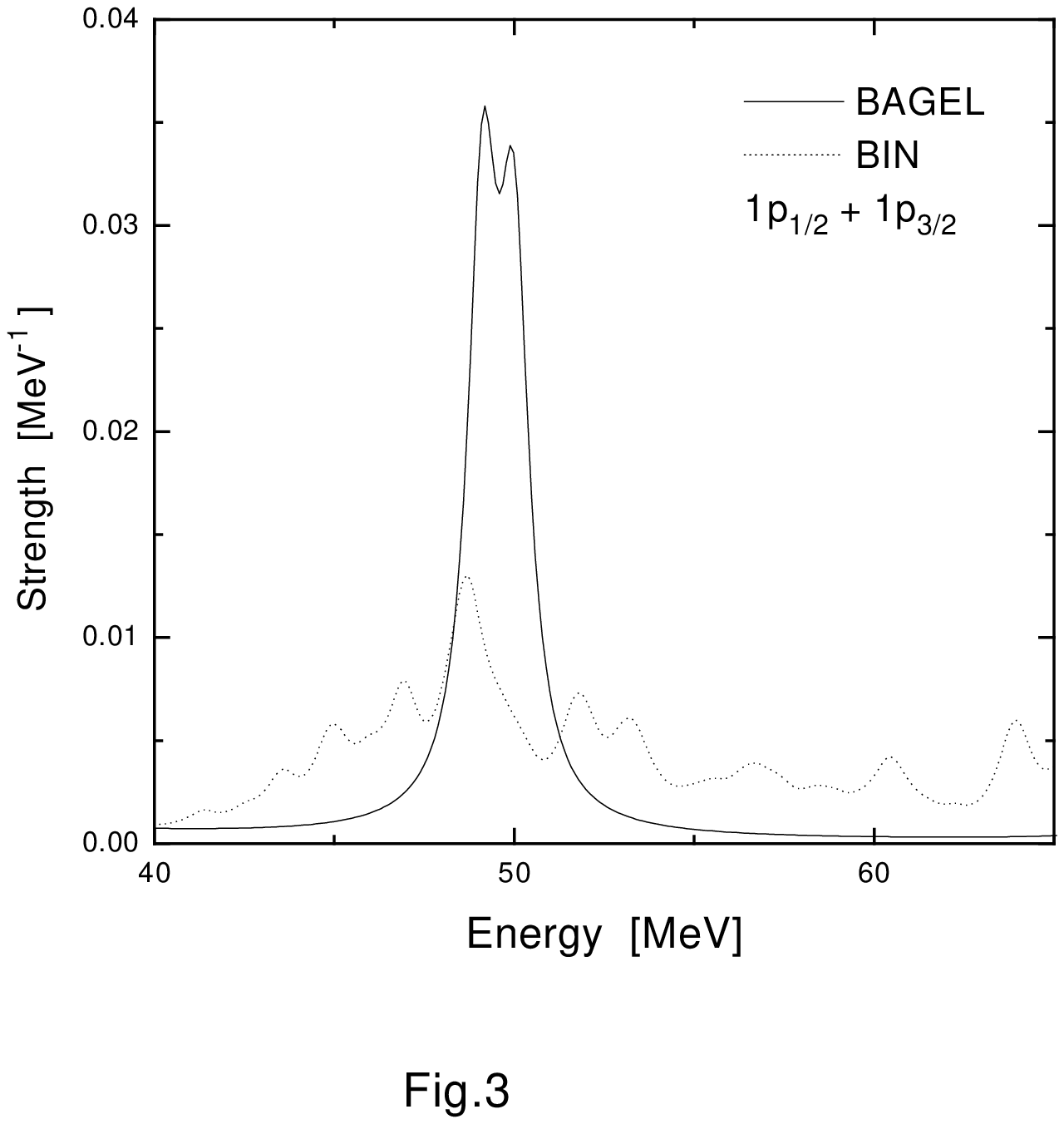}
\caption{Spectral function for $l=1$ strength in the $A=15$ system. The 
discrete spectral function has been folded with a Lorentzian having a full 
width of 1 MeV.}
\end{figure}
\newpage
\begin{table}[table1]
\caption{The HF energy $\epsilon^{HF}_{\alpha}$, RHF energy 
$\epsilon^{RHF}_{\alpha}$, 
occupation probability $n_{\alpha}$, the spectroscopic factor and position of 
the largest fragment for all s.p.\ states in $^{16}$O,  in the BAGEL and the 
BIN approximation. All energies are in MeV. All other entries are fractions 
of the degeneracy $(2j+1)$}
\begin{center}
\begin{tabular}{||c||c|c|c|c|c|c|c|c|c||}
& & BAGEL & BIN&BAGEL&BIN  &\multicolumn{2}{c|}{BAGEL} & \multicolumn{2}{c|}{BIN}\\
state & $\epsilon^{HF}_{\alpha}$ & $\epsilon^{RHF}_{\alpha}$& $\epsilon^{RHF}_{\alpha}$ 
& $n_{\alpha} (\%)$ & $n_{\alpha} (\%)$
& $E_{max}$ & $R_{max} (\%)$ & $E_{max}$ & $R_{max} (\%)$ \\ \hline 
$1s_{1/2}$ & -42.93 & -41.64 & -41.63 & 96.89 & 96.89&-45.96\, & 47.80  & -46.18\, & 37.60  \\
$1p_{3/2}$ & -20.14 & -19.38 & -19.38 & 94.99 & 94.98&-19.56\, & 88.43 & -19.53\, & 88.38  \\
$1p_{1/2}$ & -16.47 & -15.85 & -15.85 & 93.81 & 93.79& -16.18\, & 87.59 & -16.18\, & 87.54  \\
$1d_{5/2}$ & -1.295 & -0.9692 & -0.9678&2.22& 2.22& -2.075\, & 92.63 & -2.084\, & 92.61 \\
$2s_{1/2}\,$ & -0.6594\, & -0.3901 & -0.3895\,&1.89& 1.90& -1.888\, & 91.57 & -1.884\, & 91.54 \\
$1d_{3/2}$ & 4.119 & 4.284 & 4.284& 2.66 &2.66 & 2.775  & 89.05 & 2.766 & 88.98 \\
$1f_{7/2}$ & 13.65 & 13.74 & 13.74& 0.27 &0.27 & 9.616 & 63.96 & 9.366 & 56.57 \\
$2p_{3/2}$ & 10.96 & 11.22 & 11.22& 0.89 & 0.89 & 9.108 & 78.53 & 8.966 & 72.28  \\
$1f_{5/2}$ & 19.08 & 19.05 & 19.05& 0.59 & 0.59& 22.36 & 50.34 & 13.87 & 20.82  \\
$2p_{1/2}$\, & 12.66 & 12.91 & 12.91& 0.83 &0.83& 10.36 & 74.77  & 10.12 & 64.79 \\
\hline
E/A &  &&&&&  \multicolumn{2}{c|}{-6.413 MeV} &\multicolumn{2}{c|}{-6.408 MeV}\\
\end{tabular} 
\end{center}
\end{table}
\newpage
\begin{table}[table2]
\caption{Relative differences between the central moments of 
s.p.\ strength distributions obtained in the BAGEL and the 
BIN approximation, separately for the distributions in 
the $(A-1)$ system ($\Delta\mu^< $), 
in the  $(A+1)$ system ($\Delta\mu^> $), and for the total distribution 
($\Delta\mu$). }
\begin{center}
\begin{tabular}{||c|c|c|c|c|c|c||}   
   & \multicolumn {3}{c|}{$1s_{1/2}$} & \multicolumn {3}{c||}{$1d_{5/2}$\,} \\ \hline
 k &  $\Delta\mu_{k}^{<}(\%)$ &   $\Delta\mu_{k}^{>}(\%)$ &   $\Delta\mu_{k}(\%)$ & $\Delta\mu_{k}^{<}(\%)$ &   $\Delta\mu_{k}^{>}(\%)$ &   $\Delta\mu_{k}(\%)$ \\ \hline
2 & $3.096\; 10^{-3}$ & $-0.1080$ & $1.093\;10^{-3}$  & $-0.6887$ & $-4.946\; 10^{-2}$  & $-1.094\; 10^{-2}$ \\
4 & $8.008\;10^{-2}$\,  & $\;0.1066$  &$4.062\;10^{-2}$  & $-0.2463$   & $\;4.810\; 10^{-2} $  & $0.1024$ \\
6 & $0.6429$ & $2.151$ & $0.1741$ & $2.362$ & $0.4823$ & $0.5601$\\ 
8 & $3.319$ & $9.303$  & $0.7806$  & $10.27$  & $2.600$ & $2.646$  \\ 
10 \,& $11.29$ & $26.18$ & $2.909$ & $28.43$ & $9.228$ & $9.046$ \\
\end{tabular}
\end{center}
\end{table}
\end{document}